\documentclass[twocolumn]{aastex631}
\usepackage[perpage,symbol*]{footmisc}
\usepackage{color}
\usepackage{graphicx}
\usepackage{amsmath}

\newcommand{\Mpc}{\mathrm{~km~s^{-1}~Mpc^{-1}}}
\newcommand{\pn}{\gamma_{\rm{PPN}}}
\newcommand{\n}{\nonumber}
\def\({\left(}
\def\){\right)}
\def\[{\left[}
\def\]{\right]}
\newcommand{\Rmnum}[1]{\uppercase\expandafter{\romannumeral #1\relax}}
\newcommand{\rmnum}[1]{\romannumeral #1}
\shorttitle{Galaxy-Scale Test of General Relativity}
\shortauthors{Xiao-Hui Liu et al.}

\begin{document}
\title{Galaxy-Scale Test of General Relativity with Strong Gravitational Lensing}

\correspondingauthor{Jing-Zhao Qi, Xin Zhang}
\email{qijingzhao@mail.neu.edu.cn, zhangxin@mail.neu.edu.cn}

\author{Xiao-Hui Liu}
\affiliation{Department of Physics, College of Sciences, Northeastern
University, Shenyang 110819, China}

\author{Zhen-Hua Li}
\affiliation{Department of Physics, College of Sciences, Northeastern
University, Shenyang 110819, China}

\author{Jing-Zhao Qi}
\affiliation{Department of Physics, College of Sciences, Northeastern
University, Shenyang 110819, China}

\author{Xin Zhang}
\affiliation{Department of Physics, College of Sciences, Northeastern
University, Shenyang 110819, China}
\affiliation{Frontiers Science Center for Industrial Intelligence and Systems Optimization, Northeastern University, Shenyang 110819, China}
\affiliation{Key Laboratory of Data Analytics and Optimization for Smart Industry (Northeastern University), Ministry of Education, Shenyang 110819, China}

\begin{abstract}
Although general relativity (GR) has been precisely tested at the solar system scale, precise tests at a galactic or cosmological scale are still relatively insufficient. Here, in order to test GR at the galactic scale, we use the newly compiled galaxy-scale strong gravitational lensing (SGL) sample to constrain the parameter $\pn$ in the parametrized post-Newtonian (PPN) formalism. We employ the Pantheon sample of type Ia supernovae observation to calibrate the distances in the SGL systems using the Gaussian Process method, which avoids the logical problem caused by assuming a cosmological model within GR to determine the distances in the SGL sample. Furthermore, we consider three typical lens models in this work to investigate the influences of the lens mass distributions on the fitting results. We find that the choice of the lens models has a significant impact on the constraints on the PPN parameter $\pn$. By using the minimum $\chi^2$ comparison and the Bayesian information criterion as evaluation tools to make comparisons for the fitting results of the three lens models, we find that the most reliable lens model gives the result of $\pn=1.065^{+0.064}_{-0.074}$, which is in good agreement with the prediction of $\pn=1$ by GR. As far as we know, our 6.4\% constraint result is the best result so far among the recent works using the SGL method.

\end{abstract}


\section{Introduction}

It is well known that Einstein's theory of general relativity (GR), one of the cornerstones of modern physics, is an extremely successful description of gravity, which has been confirmed in all current experiments, especially in some famous experiments, e.g.  \cite{Dyson:1920cwa,Pound:1960zz,Shapiro:1964uw,Taylor:1979zz}. In recent years, the continued pursuit for tests of GR at a much higher precision has yielded many accomplishments, especially at the solar system scale. However, the long-range nature of gravity on the extragalactic scale is still poorly understood. Any possible violation of GR would have the potential to advance our understanding of many important problems in fundamental physics, for which a good example is the accelerating expansion of the Universe.

In the framework of GR, an exotic component with negative pressure, called dark energy, was introduced to explain the cosmic accelerated expansion. However, up to now, not only the nature of dark energy is still poorly understood, but also any current phenomenological models of it are more or less burdened with some theoretical problems \citep{Weinberg:1988cp,Carroll:2000fy,Copeland:2006wr}. Another perspective for understanding the cosmic acceleration is modifying Einstein's theory of gravity. Indeed, the true gravity theory may deviate GR at cosmological scale \citep{Sotiriou:2008rp}. Some geometric effects from modifying gravity could also successfully explain the cosmic accelerated expansion without introducing exotic dark energy \citep{Sotiriou:2008rp,Qi:2017xzl,Qi:2014yxa,Pan:2021tpk,Bengochea:2008gz,Zhang:2014lfa,Li:2015poa,Geng:2015hen,Zhao:2017jma}. Therefore, clarifying whether it is dark energy or modified gravity that is responsible for the accelerating expansion of the Universe requires an accurate test of GR on cosmological scales. At cosmological scales, GR with dark energy could be distinguished from modified gravity by accurately measuring the growth of structure \citep{Zhang:2007nk,Reyes:2010tr,Song:2006ej}. Moreover, the test of GR at the galactic scale could also be well established, which is the focus of this paper, and its details will be described in the following.

The validity of GR can be verified by constraining the parametrized post-Newtonian (PPN) parameter $\pn$, which describes how much space curvature is provided by the unit rest mass of the objects along or near the path of particles \citep{Thorne:1971iat,Bertotti:2003rm}. 
The PPN parameter $\pn=1$ for GR. Thus, any deviation from $\pn=1$ on large scales indicates a possible violation of GR.

Strong gravitational lensing (SGL) could provide an effective way to constrain the PPN parameter on the galactic scale. This method was first adopted in \citep{Bolton:2006yz}. They used 15 lenses sample from Sloan Lens ACS (SLACS) Survey \citep{Bolton:2005nf}, giving a result of $\pn=0.98\pm0.07$ \citep{Bolton:2006yz}, based on priors on galaxy structure taken from local observations. Subsequently, many works have provided effective tests for GR with different and larger SGL samples \citep{Smith:2009fn,Schwab:2009nz,Cao:2017nnq,collett2018precise,Yang:2018bdf}. More recently, a larger SGL sample including 161 galaxy-scale strong lensing systems has been released \citep{Chen:2018jcf}, which makes it possible to further explore the validity of GR on galactic scale at higher precision.

In this paper, we will not only improve the constraints on the PPN parameter with the larger SGL sample in exploring the validity of GR, but also refine the test of GR from two aspects compared to the previous works motioned above \citep{Bolton:2006yz,Smith:2009fn,Schwab:2009nz,Cao:2017nnq}. Firstly, it should be emphasized that the influence of lens model selection cannot be ignored in this method for testing GR, since the theoretical value of light deflection is determined by the mass distribution model of lens galaxies and gravitational theory together. In the previous works, most of them only considered a single lens mass model. We will fully consider the different lens mass models and investigate their impacts on the results. Secondly, in previous related works, the distance information required to constrain the PPN parameter is conventionally provided by the prediction of the standard $\Lambda$ cold dark matter ($\Lambda$CDM) model that is derived within the framework of GR, which more or less leads to a logic problem in testing gravity. To avoid this problem, in this work, by using the method of Gaussian Process, we will construct a smooth distance-redshift curve directly from the type \Rmnum{1}a supernovae (SN Ia) observation to calibrate the distances in the SGL sample.

This paper is organized as follows. In Section \ref{MO}, we describe the methodology of testing GR by SGL and Gaussian Process. In addition, the SGL sample and the latest Pantheon SN Ia sample are also introduced. In Section \ref{RD}, the constraint results in different lens mass models, and the model comparison based on the information criterion are presented. Finally, the conclusions and discussions are given in Section \ref{Cs}.

\section{Methodology and Observational Data} \label{MO}

In the limit of a weak gravitational field, the PPN form of Schwarzschild metric can be expressed as
\begin{equation}
\begin{aligned}
d s^{2}=& c^{2} d t^{2}\left(1-2 G M / c^{2} r\right)-d r^{2}\left(1+2 \pn G M / c^{2} r\right) \\
&-r^{2} d \Omega^{2},
\end{aligned}
\end{equation}
where $\pn$ is the PPN parameter, and $\pn=1$ stands for GR. $M$ is the mass of the central object, in this case the mass of the lens galaxy.

\subsection{Strong Gravitational Lensing Systems}

For a SGL system, the Einstein radius $\theta_{\mathrm{E}}$ reflecting the angular separations between multiple images is corresponding to the mass contained within a cylinder of Einstein radius $M_{\mathrm{E}}$,
\begin{equation}\label{TE}
\theta_{\mathrm{E}}=\sqrt{\frac{1+\pn}{2}}\left(\frac{4 G M_{\mathrm{E}}}{c^{2}} \frac{D_{\rm{d s}}}{D_{\rm{s}} D_{\rm{d}}}\right)^{1 / 2},
\end{equation}
where $D_{\rm{s}}$ is the angular diameter distance between observer and source, $D_{\rm{d}}$ is the angular diameter distance between observer and lens, and $D_{\rm{ds}}$ is the angular diameter distance between lens and source. By introducing the  radius perpendicular to the line of sight $R_{\rm{E}}=D_{\rm{d}}\theta_{\rm{E}}$ and rearranging Eq. (\ref{TE}), we can obtain 
\begin{equation}
\frac{G M_{\mathrm{E}}}{R_{\mathrm{E}}}=\frac{2}{(1+\pn)} \frac{c^{2}}{4} \frac{D_{\rm{s}}}{D_{\rm{d s}}} \theta_{\mathrm{E}}.
\end{equation}
It should be noted that the mass $M_{\mathrm{E}}$ is derived from the gravitational theory and the observations of angular separations between multiple images, and thus it could also be called the gravitational mass, denoted as $M_{\mathrm{grl}}^{\mathrm{E}}$. On the other hand, by observing the velocity dispersion of the lens galaxy and assuming a lens mass model, the dynamical mass enclosed within the Einstein radius, $M_{\mathrm{dyn}}^{\mathrm{E}}$, can also be inferred. In theory, these two masses should be equivalent, i.e., $M_{\mathrm{grl}}^{\mathrm{E}}=M_{\mathrm{dyn}}^{\mathrm{E}}$ \citep{Chen:2018jcf}.

To estimate the dynamical mass from the velocity dispersion $\sigma$ of lens galaxy, here we choose a general mass model based on power-law density profiles to describe the mass distribution of lens galaxy \citep{Koopmans:2005ig}:
\begin{equation}\label{Eq_lens}
\left\{\begin{array}{l}
\rho(r)=\rho_{0}\left(r / r_{0}\right)^{-\gamma} \\
v(r)=v_{0}\left(r / r_{0}\right)^{-\delta} \\
\beta(r)=1-\sigma_{\theta}^{2} / \sigma_{r}^{2},
\end{array}\right.
\end{equation}
where $\rho(r)$ is the total mass density distribution including luminous matter and dark matter, and $v(r)$ represents the density distribution of luminous mass. The parameter $\beta(r)$ characterizes the anisotropic distribution of velocity dispersion pattern, where $\sigma_{\theta}$ and $\sigma_{r}$ are the tangential and radial components of the velocity dispersion, respectively. Here $\gamma$ and $\delta$ are power law indices; note that $\gamma$ should not be confused with the PPN parameter $\pn$. If $\gamma=\delta=2$ and $\beta=0$, this mass model will reduce to the singular isothermal sphere (SIS) model commonly used in studies of SGL. It is important to note that the total mass density slope $\gamma$ is probably dependent on redshift \citep{Ruff:2010rv,Bolton:2012uh,Cao:2016wor,Cui:2017idf}. Therefore, three parametrized models of $\gamma$ commonly used will be considered in this work, namely
\begin{eqnarray} \nonumber
&(\rm{\rmnum{1}})& P_{1}: \gamma=\gamma_{0}, \\ 
&(\rm{\rmnum{2}})& P_{2}: \gamma=\gamma_{0}+\gamma_{z}  z_{\rm{d}}, \\ \nonumber
&(\rm{\rmnum{3}})& P_{3}: \gamma=\gamma_{0}+\gamma_{z} z_{\rm{d}}+\gamma_{s} \log \tilde{\Sigma},
\end{eqnarray}
where $\gamma_0$, $\gamma_z$ and $\gamma_s$ are arbitrary constants. For $P_1$ model, the dependences on redshift and mass distribution are not considered, and this model is also frequently used in previous related works \citep{Smith:2009fn,Schwab:2009nz,Cao:2017nnq}. $P_2$ model assumes that the total mass density distribution of galaxies evolves with redshift. $z_d$ is the redshift of lens galaxy. In $P_3$ model, in addition to the dependence of redshift, the dependence on the surface mass density in the mass distribution is also considered. Here, $\tilde{\Sigma}$ represents the normalized surface mass density, expressed as \citep{Chen:2018jcf}
\begin{equation}
\tilde{\Sigma}=\frac{\left(\sigma_{0} / 100 \mathrm{~km} \mathrm{~s}^{-1}\right)^{2}}{R_{\mathrm{eff}} / 10 h^{-1} \mathrm{kpc}},
\end{equation}
where $h=H_0/(100~\Mpc)$ and $H_0$ is the Hubble constant. $\sigma_0$ is the observed velocity dispersion, and $R_{\mathrm{eff}}$ is the half-light radius of the lens galaxy.

Based on the well-known spherical Jeans equation and the mass density profiles in Eq. (\ref{Eq_lens}), the mass enclosed within a sphere with radius $r$ is \citep{Schwab:2009nz}
\begin{equation} \label{Mr}
M(r)=\frac{2}{\sqrt{\pi}} \frac{\Gamma(\gamma / 2)}{\Gamma\left(\frac{\gamma-1}{2}\right)}\left(\frac{r}{R_{\mathrm{E}}}\right)^{3-\gamma} M_{\mathrm{dyn}}^{\mathrm{E}},
\end{equation}
where $\Gamma(x)$ is Euler's Gamma function. The radial velocity distance dispersion of the luminous mass $\sigma_r$  could be expressed as
\begin{equation} \label{sigma}
\sigma_{r}^{2}(r)=\frac{G \int_{r}^{\infty} \mathrm{d} r^{\prime} r^{\prime 2 \beta-2} v\left(r^{\prime}\right) M\left(r^{\prime}\right)}{r^{2 \beta} v(r)}.
\end{equation}
By taking $\xi=\gamma+\delta-2$ and assuming $\beta$ is independent of the radius $r$, the combination of Eqs. (\ref{Mr}) and (\ref{sigma}) could infer the expression of $\sigma_r^2$, 
\begin{equation} \label{sigma_r}
\sigma_{r}^{2}(r)=\frac{2}{\sqrt{\pi}} \frac{G M_{\mathrm{dyn}}^{\mathrm{E}}}{R_{\mathrm{E}}} \frac{1}{\xi-2 \beta} \frac{\Gamma(\gamma / 2)}{\Gamma\left(\frac{\gamma-1}{2}\right)}\left(\frac{r}{R_{\mathrm{E}}}\right)^{2-\gamma}.
\end{equation}
In the observation of SGL, the actual measurement of velocity dispersion is the component of luminosity-weighted along the line of sight (LOS) and over the effective spectrometer aperture. Moreover, for the spectroscopic data, the velocity dispersion $\sigma_{\rm{ap}}$ measured within certain apertures $\theta_{\rm{ap}}$ should be normalized to the one within typical physical aperture via $\sigma_{0}=\sigma_{\mathrm{ap}}\left[\theta_{\mathrm{eff}} /\left(2 \theta_{\mathrm{ap}}\right)\right]^{\eta}$, where $\theta_{\mathrm{eff}}$ is the effective angular radius of the lens galaxy, and the value of the correction factor $\eta$ we adopt is $\eta=-0.066\pm0.033$ \citep{Jorgensen:1995zz}. The detailed derivation and description could be found in \cite{Koopmans:2005ig,Chen:2018jcf}. Taking all of these into account,  the theoretical value of the velocity dispersion could be expressed as \citep{Chen:2018jcf}
\begin{equation}\label{sigma0}
\sigma_{0}=\sqrt{\frac{c^{2}}{2 \sqrt{\pi}} \frac{2}{(1+\pn)} \frac{D_{\mathrm{s}}}{D_{\mathrm{ds}}} \theta_{\mathrm{E}} f(\gamma, \delta, \beta)\left(\frac{\theta_{\mathrm{eff}}}{2 \theta_{\mathrm{E}}}\right)^{(2-\gamma)}},
\end{equation}
where
\begin{equation}
\begin{aligned}
f(\gamma, \delta, \beta)=& \frac{(2 \sqrt{\pi})(3-\delta)}{(\xi-2 \beta)(3-\xi)} \\
& \times\left[\frac{\Gamma[(\xi-1) / 2]}{\Gamma(\xi / 2)}-\beta \frac{\Gamma[(\xi+1) / 2]}{\Gamma[(\xi+2) / 2]}\right] \\
& \times \frac{\Gamma(\gamma / 2) \Gamma(\delta / 2)}{\Gamma[(\gamma-1) / 2] \Gamma[(\delta-1) / 2]}.
\end{aligned}
\end{equation}
Considering the extra mass contribution from matters along the LOS, the fractional uncertainty of velocity dispersion at the level of ~3\% will be taken as the systematic error \citep{Jiang:2007jx}, and the total uncertainty of $\sigma_0$ from the observation is 
\begin{eqnarray} \label{sigma_SGL}
\left(\Delta \sigma_{0}\right)^{2}&=\left(\Delta \sigma_{0}^{\mathrm{stat}}\right)^{2}+\left(\Delta \sigma_{0}^{\mathrm{AC}}\right)^{2}+\left(\Delta \sigma_{0}^{\mathrm{sys}}\right)^{2}\\ \nonumber
=&\left[\frac{\Delta \sigma_{\rm{ap}}^2}{\sigma_{\rm{ap}}^2}+0.03^2+\left(\ln\left(\frac{\theta_{\rm{eff}}}{2\theta_{\rm{ap}}}\right)\Delta\eta\right)^2\right]\sigma_0^2,
\end{eqnarray}
where $\Delta \sigma_{0}^{\mathrm{stat}}$ is the statistical uncertainty propagated from the measurement error of $\sigma_{\rm{ap}}$, and $\Delta \sigma_{0}^{\mathrm{AC}}$ is caused by the uncertainty of $\eta$.

\subsection{SGL Data Sample}
The SGL sample we use in this paper includes 161 galaxy-scale SGL systems compiled by \citet{Chen:2018jcf} from the following surveys: the Lens Structure and Dynamics (LSD) survey \citep{koopmans2002stellar,Koopmans:2002qh,Treu:2002ee,Treu:2004wt}, the Strong Lensing Legacy Survey \citep{Ruff:2010rv,Sonnenfeld:2013xga,Sonnenfeld:2014gpa}, the Sloan Lens ACS survey \citep{Bolton:2008xf,Auger:2009hj,Auger:2010va,Shu:2014aba,Shu:2017yon}, and the Baryon Oscillation Spectroscopic Survey (BOSS) emission-line lens survey \citep{Brownstein:2011leg,Shu_2016,shu2016boss}, which are selected with strict criteria to ensure the validity of the assumption of spherical symmetry on the lens galaxy. Specifically, the lens galaxies should be early-type galaxies with E/S0 morphologies and should not have significant substructures or close massive companions. The information of these selected 161 SGL systems could be found in Appendix of \citet{Chen:2018jcf}.

As mentioned above, the power-law density profiles expressed by Eq. (\ref{Eq_lens}) could describe the mass distribution of lens galaxy well, in which the slope of the luminosity density profile $\delta$ for a specific lens galaxy can be obtained by observation actually, although it is commonly treated as a universal parameter for all lens galaxies. By fitting the two-dimensional power-law luminosity profile over a circle of radius $\theta_{\rm{eff}}/2$ for lens galaxies with the high-resolution \textit{Hubble Space Telescope} imaging data, the measured values of $\delta$ for the 130 lens galaxies in the full sample were obtained \citep{Chen:2018jcf}. By constraining cosmological parameters with the full sample ($\delta$ as a universal parameter) and the truncated sample ($\delta$ as an observable for each SGL system), respectively, \citet{Chen:2018jcf} concluded that the intrinsic scatter of $\delta$ among the lenses should be taken into account to get an unbiased cosmological estimate. For the detailed analysis and descriptions, one can refer to \citet{Chen:2018jcf} and references therein.

In this paper, therefore, the SGL sample we adopt is the truncated sample including 130 SGL systems with the intrinsic scatter of $\delta$, for which the redshift range of lenses is $0.0624 \leq z_{\rm{d}} \leq 0.7224$ and the redshift range of sources is $0.1970 \leq z_{\rm{s}} \leq 2.8324$. However, in the following, we will use the Gaussian process to reconstruct a distance-redshift curve directly from the SN Ia data and then to calibrate the distances of the SGL systems. Therefore, the maximum redshift of the SN Ia sample determines the number of SGL systems that can be calibrated. Actually, only 120 SGL data are available in this paper.

\subsection{Gaussian Process}

To constrain on the PPN parameter $\pn$, it is necessary to determine the distances $D_{\rm{d}}$ and $D_{\rm{ds}}$, which are conventionally derived from the $\Lambda$CDM model \citep{Cao:2017nnq,Schwab:2009nz}. However, theoretically, the logic of this approach is problematic because the $\Lambda$CDM is derived within the framework of GR. Therefore, in this paper, we will calibrate the distances of SGL via a cosmological model-independent method, Gaussian Process (GP), which could make us reconstruct a smooth distance-redshift curve from SN \Rmnum{1}a observation without assuming any parametrization forms. This reconstruction method has been widely used in cosmology \citep{Wang:2019yob,Qi:2018aio,Seikel:2012uu,Seikel:2012cs,Wang:2020dbt,liao2019model}.

In GP, the values of the reconstructed function at different points $z$ and $\tilde{z}$ are connected by a covariance function $k(z, \tilde{z})$. According to the analysis in \cite{Seikel:2013fda}, the squared exponential form with the Mat\'{e}rn $(\nu=9/2)$ covariance function can lead to more reliable results than all others. So we take it here, and its expression is
\begin{eqnarray}
k(z,\tilde{z})&=&\sigma_f^2\exp(-\frac{3|z-\tilde{z}|}{\ell})\nonumber\\
&\times &(1+\frac{3|z-\tilde{z}|}{\ell}+\frac{27(z-\tilde{z})^2}{7\ell^2}\nonumber\\
&+&\frac{18|z-\tilde{z}|^3}{7\ell^3}+\frac{27(z-\tilde{z})^4}{35\ell^4}),\label{7}
\end{eqnarray}
where $\sigma_f$ and $\ell$ are hyperparameters which can be optimized by the GP itself via the observational data.

\begin{figure}
\centering 
\includegraphics[scale=0.62]{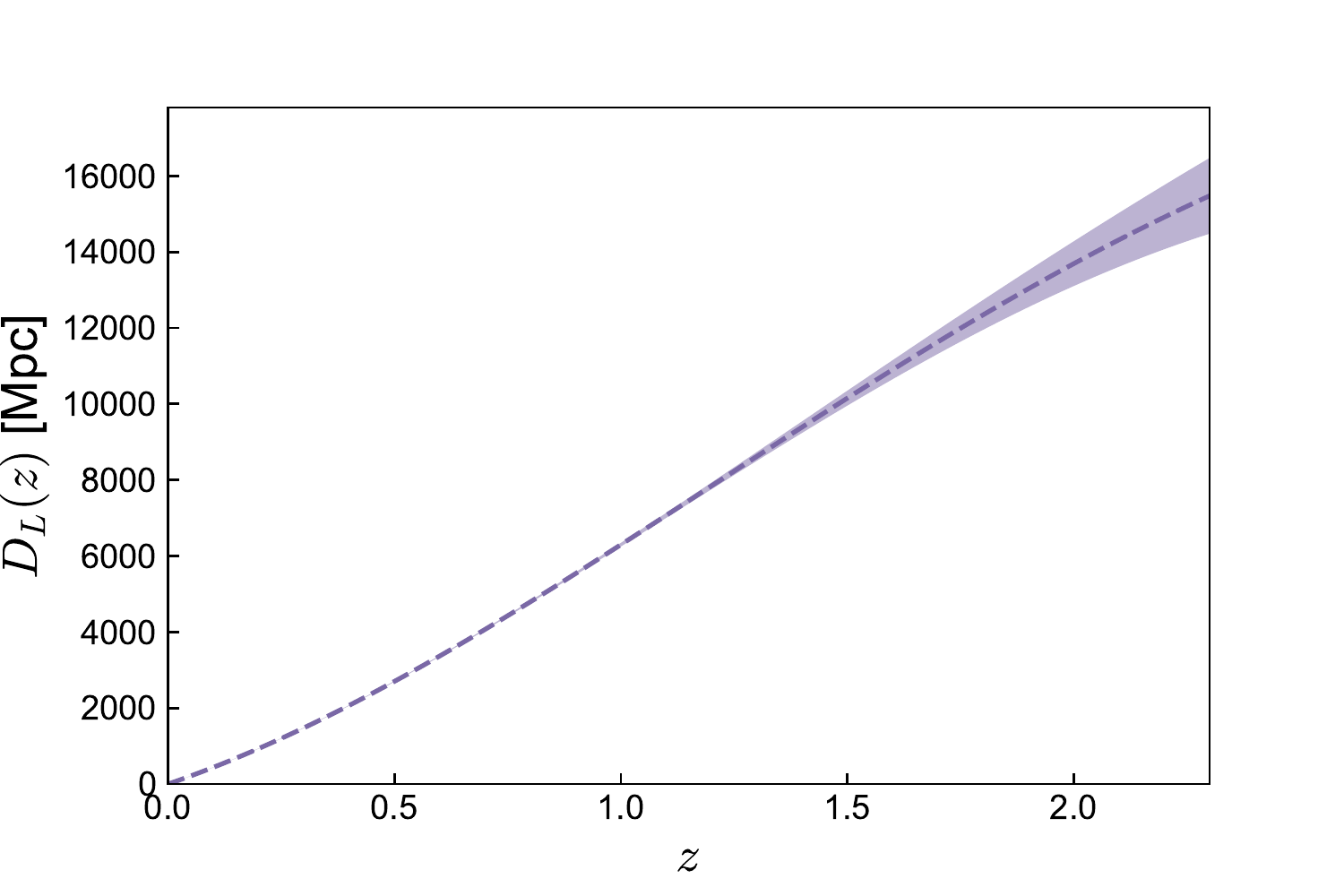}
\caption{Reconstruction of the luminosity distance $D_L(z)$ from the Pantheon sample of SN Ia observation using the Gaussian Process method. Here, the purple dashed line represents the best fit and the shaded area represents the 68\% confidence level. }\label{GPDL}
\end{figure}

\subsection{Type \Rmnum{1}a Supernovae}

The SN Ia sample we choose is the Pantheon sample from the Pan-STARRS1 Medium Deep Survey \citep{Scolnic:2017caz}, which consists of 1048 SN \Rmnum{1}a data covering redshift range $0.001<z<2.3$. This means that only the systems with $z_{\rm{s}}<2.3$ in the SGL sample can be calibrated by SN Ia. Thus, only 120 SGL systems are actually available.

For SN Ia as standard candle, the observational data are connected with the distance modulus $\mu_{\rm{obs}}$ by the SALT2 light curve fitter \citep{SNLS:2010pgl},
\begin{equation}
\mu_{\mathrm{obs}}(z)=m_{B}(z)+\alpha_{\rm{SN}} \cdot X_{1}-\beta_{\rm{SN}} \cdot \mathcal{C}-M_{B},
\end{equation}
where $m_{\rm{B}}$ represents the rest frame \textit{B}-band peak magnitude, $ X_{1}$ and $\mathcal{C}$ denote the time stretch of light curve and the supernova color at maximum brightness, respectively. $\alpha_{\rm{SN}}$ and $\beta_{\rm{SN}}$ are two nuisance parameters. The parameter $M_{B}$ is the absolute \textit{B}-band magnitude, which is exactly degenerate with $H_0$. Theoretically, if the value of $M_{B}$ or $H_0$ is determined, absolute luminosity distance can be obtained from SN Ia. In the context of Hubble tension \citep{Riess:2019cxk,DiValentino:2021izs,Zhang:2019ylr,Guo:2018ans,Guo:2018uic,Guo:2019dui,Feng:2019jqa,Gao:2021xnk,Wang:2021kxc,Dainotti:2021pqg}, the selection of $H_0$ often has an effect on the results \citep{Qi:2020rmm}. However, in this paper, the velocity dispersion we focus is just sensitive to the distance ratio as seen from Eq. (\ref{sigma0}), which eliminates the influence of the choice for $H_0$.

The stretch-luminosity parameter $\alpha_{\rm{SN}}$ and the color-luminosity parameter $\beta_{\rm{SN}}$ could be calibrated to zero through a method called BEAMS with bias corrections \citep{Kessler:2016uwi}. The expression of the observed distance modulus can be simplified as
\begin{equation}
\mu_{\mathrm{obs}}(z)=m_{B}(z)-M_{B}.
\end{equation}
The theoretical distance modulus is defined as
\begin{equation}
\mu_{\mathrm{th}}=5 \log _{10}\left(D_{L} / \mathrm{Mpc}\right)+25.
\end{equation}
The reconstruction of the luminosity distance from the Pantheon sample by using GP is shown in Figure \ref{GPDL}.

\subsection{The Likelihood Function}
By using the \texttt{emcee} Python module \citep{foreman2013emcee} based on the Markov Chain Monte Carlo (MCMC) method to maximize the likelihood $\mathcal{L} \propto e^{-\chi^{2} / 2}$, the PPN parameter ($\pn$) and the parameters of lens models ($\gamma_0$, $\gamma_z$ and $\gamma_s$) are fitted simultaneously. Specifically, the 
$\chi^2$ function is defined as
\begin{equation}\label{chi2}
\chi^{2}\left(\boldsymbol{p}, \pn \right)=\sum_{i=1}^{120} \frac{\left(\sigma_0^{\mathrm{th}}\left(z_{i} ; \pn\right)-\sigma_0^{\mathrm{obs}}\left(z_{i} ; \boldsymbol{p}\right)\right)^{2}}{\Delta\sigma_{0}^{\rm{tot}}\left(z_{i}\right)^{2}},
\end{equation}
where $\boldsymbol{p}$ represents the parameters of lens models. It should be noted that the uncertainty $\Delta\sigma^{\rm{tot}}_{0}$ is contributed by both uncertainties from the observation, $\Delta\sigma^{\rm{SGL}}_0$ given by Eq. (\ref{sigma_SGL}), and from the distance calibration by SN Ia ($\Delta\sigma^{\rm{SN}}_0$), and thus we have
\begin{equation}\label{error}
\left(\Delta\sigma^{\rm{tot}}_{0}\right)^2=\left(\Delta\sigma^{\rm{SGL}}_0\right)^2+\left(\Delta\sigma^{\rm{SN}}_0\right)^2.
\end{equation}
By using Eq. (\ref{sigma0}), the fractional uncertainty of $\sigma_0$ from the distance calibration by SN Ia could be estimated as 
\begin{equation}
\Delta\sigma^{\rm{SN}}_0= \sigma_0^{\rm{th}}\frac{\Delta D_{\rm{ratio}}}{2D_{\rm{ratio}}},
\end{equation}
where $D_{\rm{ratio}}$ is a convenient notation for the ratio of distances in Eq. (\ref{sigma0}), i.e., $D_{\rm{ratio}} \equiv D_{\rm{s}}/D_{\rm{ds}}$, and its uncertainty is $\Delta D_{\rm{ratio}}$. Through the smooth luminosity distance-redshift curve reconstructed from the SN Ia data, the angular diameter distance values of $D_{\rm{d}}$, $D_{\rm{s}}$ and their uncertainties ($\Delta D_{\rm{d}}$, $\Delta D_{\rm{s}}$) of each SGL system can be obtained by using the distance duality relation \citep{etherington1933lx,Qi:2019spg,Liao:2015uzb}. The ratio of distances $\Delta D_{\rm{ratio}}$ expressed by luminosity distances with superscript $L$ is
\begin{equation}
D_{\rm{ratio}}=\frac{1}{1-\frac{D_{\rm{d}}^L\left(1+z_{\rm{d}}\right)}{D_{\rm{s}}^L\left(1+z_{\rm{s}}\right)}},
\end{equation}
and its uncertainty is given by
\begin{eqnarray}
\Delta D_{\rm{ratio}}^2=&D_{\rm{ratio}}^4 \left(\frac{(1+z_{\rm{d}})}{D_{\rm{s}}^L\left(1+z_{\rm{s}}\right)}\Delta D_{\rm{d}}^L\right)^2\nonumber\\
&+D_{\rm{ratio}}^4\left(\frac{D_{\rm{d}}^L(1+z_{\rm{d}})}{(D_{\rm{s}}^L)^2(1+z_{\rm{s}})} \Delta D_{\rm{s}}^L\right)^2.
\end{eqnarray}

In addition, based on the previous studies \citep{Gerhard:2000ck,Bolton:2006yz,Schwab:2009nz,Cao:2017nnq,Chen:2018jcf}, we will marginalize the anisotropy parameter $\beta$ using a Gaussian prior with $\beta=0.18\pm0.13$ over the range of $\left[\bar{\beta}-2 \sigma_{\beta}, \bar{\beta}+2 \sigma_{\beta}\right]$, where $\bar{\beta}=0.18$ and $\sigma_{\beta}=0.13$.

\section{Results and Discussion}\label{RD}

\begin{table*} \label{tab_results} 
\caption{Constraint results (68.3\% confidence level) for $\pn$ and the SGL parameters ($\gamma_0$, $\gamma_z$, $\gamma_{\mathrm{s}}$) in different lens models.}
\begin{center}
\setlength{\tabcolsep}{3mm}{
\begin{tabular}{lcccccc}
\hline
\hline
Lens model & $\pn$ &$\gamma_0$ & $\gamma_z$ & $\gamma_{\mathrm{s}}$ & $\chi^2_{\rm{min}}$ &BIC \\ 
\hline
$P_{1}: \gamma=\gamma_0$ & $1.179_{-0.096}^{+0.081}$ & $2.031\pm{0.027}$ & - & - & 334.896& 349.259\\
$P_{2}: \gamma=\gamma_0+\gamma_zz_d$ & $1.455_{-0.127}^{+0.154}$ & $2.241\pm{0.039}$ & $-0.629\pm{0.101}$ & - & 292.534 & 311.684 \\
$P_{3}: \gamma=\gamma_{0}+\gamma_{z} z_{d}+\gamma_{\mathrm{s}}\ln\tilde{\Sigma}$ & $1.065_{-0.074}^{+0.064}$ & $1.265\pm{0.103}$ & $-0.246\pm{0.091}$ & $0.281\pm{0.028}$ & 186.196 & 210.133\\
\hline
$P_{3}$ in $\Lambda$CDM & $1.077_{-0.080}^{+0.061}$ & $1.293\pm0.098$ & $-0.247\pm0.077$ & $0.272\pm0.026$ & 188.742 & 213.080 \\
\hline
\hline
\end{tabular}}
\end{center}
\end{table*}

\begin{figure}
\centering 
\includegraphics[scale=0.35]{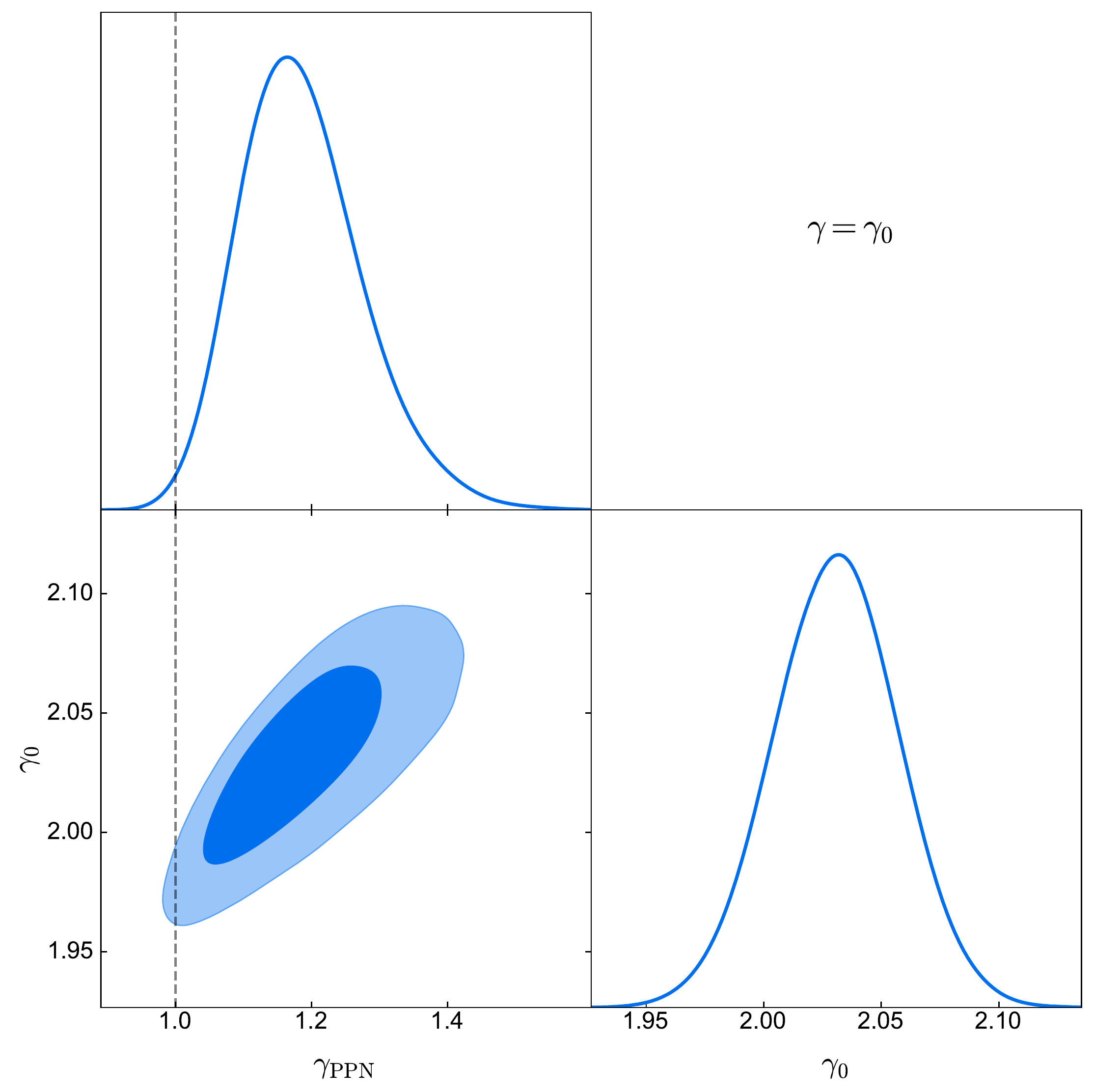}
\caption{1D and 2D marginalized probability distributions at 68.3\% and 95.4\% confidence level for the PPN parameter $\pn$ and the lens model parameter $\gamma_0$ in the $P_1$ model. The dashed line represents $\pn=1$ predicted by GR.}\label{GP1}
\end{figure}

\begin{figure*}
\centering 
\includegraphics[scale=0.5]{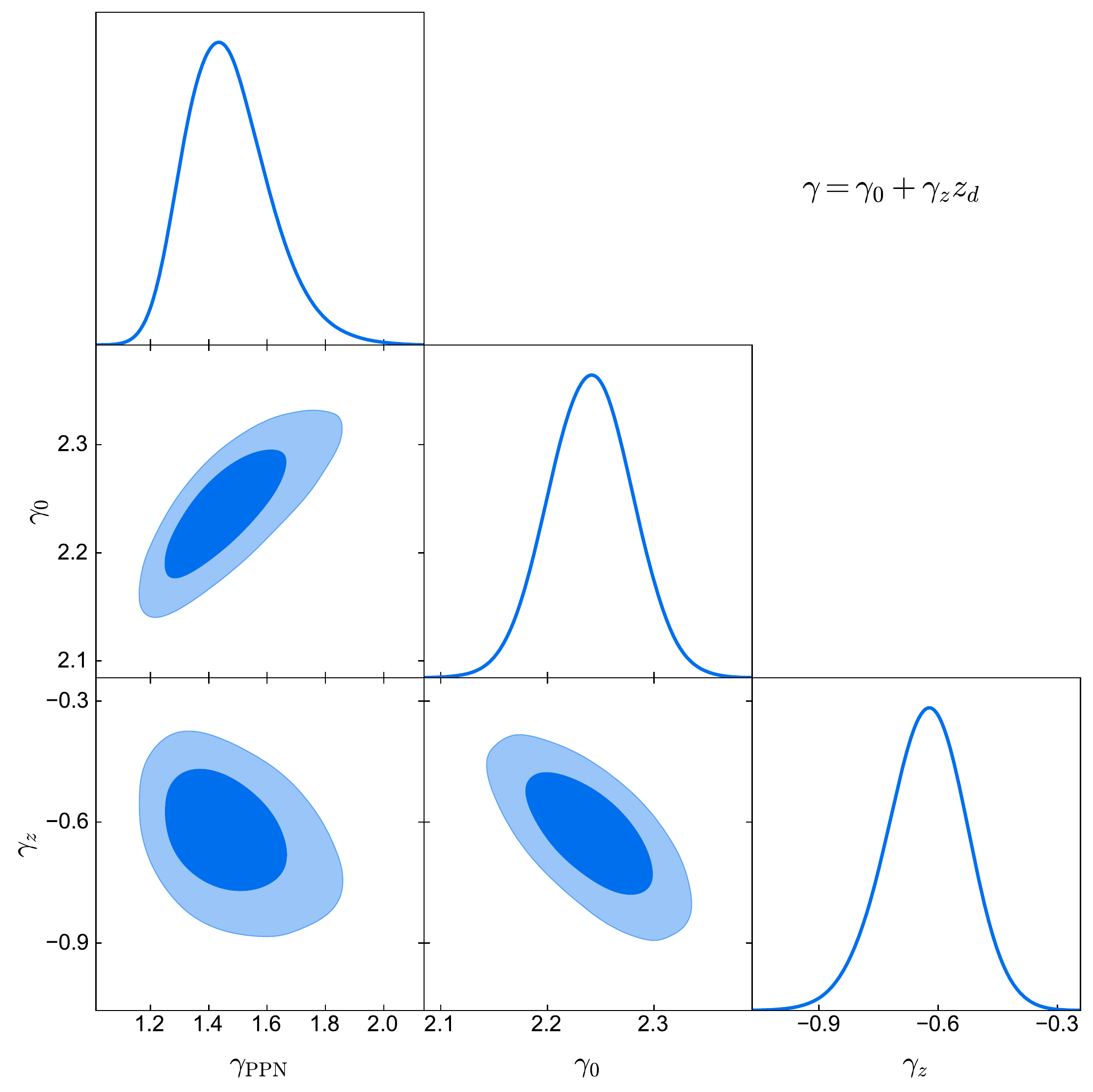}
\caption{1D and 2D marginalized probability distributions at 68.3\% and 95.4\% confidence level for the PPN parameter $\pn$ and the lens model parameters ($\gamma_0$, $\gamma_z$) in the $P_2$ model.}\label{GP2}
\end{figure*}

\begin{figure*}
\centering
\includegraphics[scale=0.5]{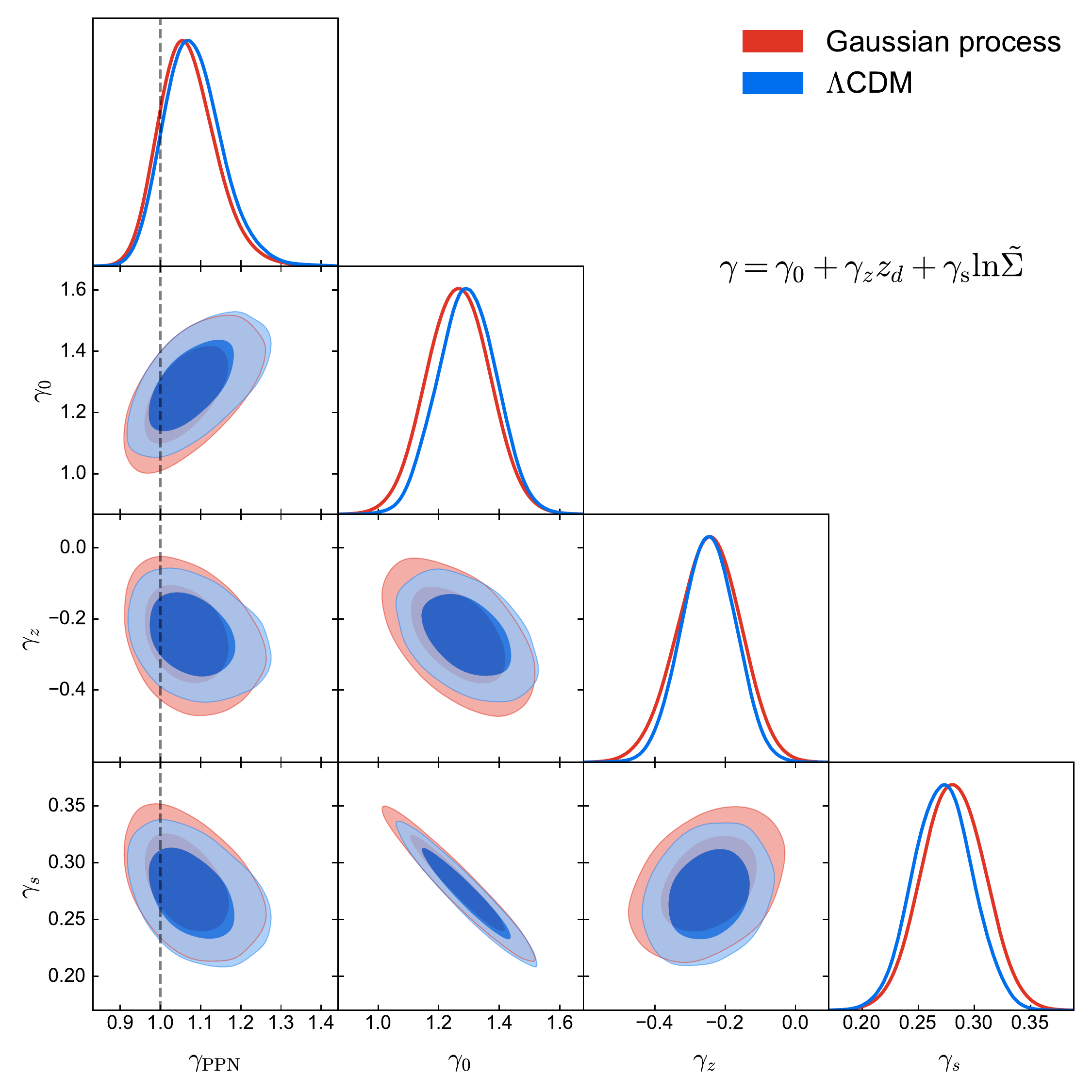}
\caption{1D and 2D marginalized probability distributions at 68.3\% and 95.4\% confidence level for the PPN parameter $\pn$ and the lens model parameters ($\gamma_0$, $\gamma_z$, $\gamma_{\mathrm{s}}$) in the $P_3$ model based on two methods of calibrating distances using the $\Lambda$CDM model and the Gaussian process, respectively. The dashed line represents $\pn=1$ predicted by GR.} \label{GP3}
\end{figure*} 

The one-dimensional and two-dimensional marginalized probability distributions of the PPN parameter and SGL parameters in three lens models are shown in Figures \ref{GP1}-\ref{GP3}, and the fitting values of parameters are presented in Table \ref{tab_results}.

For $P_1$ model, according to the constraint result, $\pn=1.179_{-0.096}^{+0.081}$, we find that the inferred value of the PPN parameter deviates from $\pn=1$ at 1$\sigma$ confidence level, but is still consistent with $\pn=1$ at 2$\sigma$ confidence level, as shown in Figure \ref{GP1}, which means that there is no significant signal indicating the violation of GR. The constraint accuracy of $\pn$ is about 7.5\%. For the SGL parameter $\gamma_0$ reflecting the total mass density distribution of lens galaxy, the constraint result of $\gamma_0=2.031\pm{0.027}$ is consistent with a spherically symmetric total mass distribution of lens galaxy ($\gamma=2$) which is predicted by the SIS model.

In Figure \ref{GP2}, we show the constraint results for $P_2$ model. It can be seen clearly that the change of the lens model has a significant impact on the results. The constraint result of the PPN parameter is $\pn=1.455^{+0.154}_{-0.127}$ indicating that the GR is excluded above 2$\sigma$ confidence level. Moreover, the constraint accuracy of $\pn$ is also reduced to 9.6\%. This reduction should not be caused alone by the addition of one parameter compared to $P_1$ model. Because from the results of $P_3$ model with additional three parameters in Table \ref{tab_results}, the constraint accuracy of $\pn$ becomes improved instead. Therefore, the reduction in constraint accuracy of $\pn$ here is likely to be caused by the lens model, which will be discussed in detail in the later analysis of the Bayesian information criterion. Concerning the parameters of the lens model, the constraint result of $\gamma_0=2.241\pm{0.039}$ indicates that the total mass density distribution of lens galaxy is not spherically symmetric. In addition, the parameter $\gamma_z$ represents the dependence of the total density profile of lens galaxy on the redshift. From the result $\gamma_z=-0.629\pm{0.101}$, we find that the mass distribution of lens galaxies, usually early-type galaxies, is likely evolving with cosmic time.

For $P_3$ model, as shown in Figure \ref{GP3} and Table \ref{tab_results}, the inferred value of the PPN parameter with $\pn=1.065_{-0.074}^{+0.064}$ is in good agreement with the prediction of $\pn=1$ by GR, and its constraint accuracy is improved to about 6.4\%, although more parameters are considered in the fitting. For the parameters of lens model, the fit value of $\gamma_0=1.265\pm{0.103}$ is much smaller than those of the above two lens models, consistent with the previous results of \citet{Chen:2018jcf} obtained by constraining $\Lambda$CDM model with the same SGL sample used here. The other two parameters of lens model are constrained to be $\gamma_z=-0.246\pm{0.091}$ and $\gamma_{\mathrm{s}}=0.281\pm{0.028}$, both excluding zero at above the 2$\sigma$ confidence level, indicating that the dependence of the total density profile on redshift and surface mass density cannot be ignored.

As we can see, the choice of lens model leads to rather different results for constraints on the PPN parameter $\pn$, which urges us to explore which models are better matched to the observational data. In general, the minimum $\chi^2$ comparison is a sufficient tool for comparing different models with the same number of parameters. A smaller value of $\chi^2_{\rm{min}}$ means a favored model by observations. In Table \ref{tab_results}, the corresponding $\chi^2_{\rm{min}}$ for each lens model is listed. However, in the case of models with different number of parameters, more parameters usually result in a smaller $\chi^2_{\rm{min}}$ for the model. To be fair, in this work, we also consider using the Bayesian information criterion (BIC) \citep{Schwarz:1978tpv} as an evaluation tool to compare different models. The BIC is expressed as
\begin{equation}
\mathrm{BIC}=-2 \ln \mathcal{L}_{\max }+k \ln N,
\end{equation}
where $\mathcal{L}_{\max }$ is the maximum likelihood, $k$ is the number of model parameters, and $N$ is the number of data points. A model with a smaller value of BIC is believed to be more favored by observations. In the model comparison, the absolute value of BIC is not important, and what we are truly interested in are the relative values of BIC between different models. Usually, $\Delta\rm{BIC}=2$ is regarded as positive evidence, and $\Delta\rm{BIC}\geq 6$ is considered to be strong evidence against the model with higher value of BIC \citep{Mukherjee:1998wp,Liddle:2004nh}. The corresponding BIC value for each lens model is listed in Table \ref{tab_results}. It can be seen that the BIC and $\chi^2_{\rm{min}}$ values of $P_3$ model are much smaller than those of the other two models, which strongly indicates that $P_3$ model is the most compatible lens model with the observation.

On the other hand, as mentioned above, to avoid the logical problem caused by assuming a cosmological model within GR adopted in the previous works commonly, we use the GP method with the latest Pantheon SN Ia sample to calibrate the distances of the SGL systems. Here, we wish to investigate whether using a cosmological model to calibrate distances influences the inference of $\pn$. For the comparison, in Figure \ref{GP3} and Table \ref{tab_results}, we present the constraint results of $P_3$ model assuming a flat $\Lambda$CDM model with parameters taken from the \textit{Planck} 2018 results \citep{Aghanim:2018eyx}. It could be seen clearly that the inferred value of the PPN parameter with $\pn=1.077_{-0.080}^{+0.061}$ is also consistent with $\pn=1$ at $1\sigma$ confidence level. Compared with the results obtained by the GP method, the best-fit value of $\pn$ has a very slight change, and their constraint precisions are almost the same order, which means that the method of calibrating distances by the cosmological model does not have a substantial impact on the inference of $\pn$. For the constraint results of lens model parameters in the framework of $\Lambda$CDM model, their best-fit values have few changes compared to the results obtained by the GP method, and their constraint precision is slightly improved because the distances provided by the cosmological model have no error, whereas the distances provided by the GP method contain. Specifically, in Eq. (\ref{error}), the term of $\left(\Delta\sigma^{\rm{SN}}_0\right)$ from the distance calibration by SN Ia is absent in the framework of $\Lambda$CDM model. On the other hand, this also leads to a larger denominator of Eq. (\ref{chi2}) in the fitting process for this method, resulting in larger $\chi^2_{\rm{min}}$ and BIC values compared to that obtained by GP method.

\section{Conclusion} \label{Cs}

Strong gravitational lensing provides us with an effective way to test GR by constraining the PPN parameter at the galactic scale. In this paper, we consider to improve the constraints on the PPN parameter in the following three aspects. (\rmnum{1}) A larger SGL sample obtained using high-resolution imaging and stellar dynamical data is considered, in which the luminosity density slope $\delta$ for each lens is treated as an observational quantity rather than a universal parameter for all lenses. (\rmnum{2}) To determine the distances in the SGL sample, we use the GP method to construct a smooth distance-redshift curve directly from the latest Pantheon SN Ia sample. 
The logical problem caused by the introduction of cosmological model in GR can thus be avoided in such a way of from observation to observation. (\rmnum{3}) The construction of lens mass model has a great influence on the constraints on the PPN parameter. Here we consider three commonly used lens models and investigate their impacts on the fit results. The most favored lens model can be selected by a model comparison using BIC and $\chi^2_{\rm{min}}$.

For the three lens models, we obtain rather different constraint results for the PPN parameter. First, for $P_1$ model, the inferred value of $\pn$ deviates from $\pn=1$ at 1$\sigma$ confidence, but the result is still consistent with GR at 2$\sigma$ confidence level. Next, for $P_2$ model, GR is excluded at above 2$\sigma$ confidence level. Finally, for $P_3$ model, the constraint result of $\pn=1.065_{-0.074}^{+0.064}$ is in good agreement with the prediction of GR. Therefore, we conclude that the choice of lens models has a great influence on the constraints on the PPN parameter as expected. 

Better characterization and modeling for the mass-dynamical structure of lensing galaxies are necessary for strong gravitational lensing to be used as an important probe to test the validity of GR. The ongoing and future massive surveys with high-quality imaging and spectroscopy such as the Dark Energy Survey \citep{frieman2013dark}, the \textit{Euclid} Survey \citep{Euclid:2021osj}, the Wide Field Infrared Survey Telescope \citep{Spergel:2015sza}, and the Large Synoptic Survey Telescope \citep{LSSTScience:2009jmu} will be very helpful in this study.

To make a model selection by comparing the three lens models, we use the BIC to assess the fitting results. We find that the BIC value of the $P_3$ model is much smaller than those of the other two models, and thus we conclude that the $P_3$ model is the most compatible model with the observation among the three lens models, which is also confirmed by another indicator of comparing the models, $\chi^2_{\rm{min}}$.

In previous related works, the distances of SGL systems are generally calibrated assuming a specific cosmological model within GR, which more or less leads to a logic problem in testing GR. To investigate the impact of this problem on the inference of $\pn$, we compare the results in the $P_3$ model based on two methods of calibrating distance using the $\Lambda$CDM model and the GP method, and we find that using a cosmological model within GR to calibrate the distances does not have a significant impact on inferring $\pn$.

The result of $\pn=1.065_{-0.074}^{+0.064}$ in the $P_3$ model is at a 6.4\% accuracy, which is the most precise constraint result obtained so far among the recent tests on GR using the similar method. Here we make a comparison with some recent representative results. 
For instance, considering uncertainties of total mass profile, velocity anisotropy, and luminous matter profile as the same in this work, but using a different SGL sample, \citet{Cao:2017nnq} obtained the constraint accuracy of the PPN parameter to be ~25\%. The nearby SGL system, ESO 325-G004, provided a constraint on the PPN parameter, $\pn=0.97\pm0.09$ at 68\% confidence level, with a 9.3\% accuracy \citep{collett2018precise}. Using four time-delay measurements of SGL from $H_0$ Lenses in COSMOGRAIL’s Wellspring collaboration, \citet{Yang:2020eoh} obtained a constraint with an accuracy of 20\%, and they forecasted that the joint constraint from 40 time-delay strong lensing could reach the accuracy of 7.7\% for the PPN parameter in the next few years. Our results show that SGL could be used as an effective tool to test GR at the galactic scale, and more promising studies can be expected with the help of more large survey telescopes in the future.

\section*{Acknowledgments}
We thank Yun Chen for very helpful discussions and for sharing the SGL data with us. {We are grateful to the referee for useful comments.} This work was supported by the National Natural Science Foundation of China (Grants Nos. 11975072, 11835009, and 11875102), the Liaoning Revitalization Talents Program (Grant No. XLYC1905011), the Fundamental Research Funds for the Central Universities (Grant Nos. N2005030 and N2105014), the National 111 Project of China (Grant No. B16009), and the China Manned Space Project (Grant No. CMS-CSST-2021-B01).

\bibliography{HK_ref}

\begin{thebibliography}{}
\expandafter\ifx\csname natexlab\endcsname\relax\def\natexlab#1{#1}\fi
\providecommand{\url}[1]{\href{#1}{#1}}
\providecommand{\dodoi}[1]{doi:~\href{http://doi.org/#1}{\nolinkurl{#1}}}
\providecommand{\doeprint}[1]{\href{http://ascl.net/#1}{\nolinkurl{http://ascl.net/#1}}}
\providecommand{\doarXiv}[1]{\href{https://arxiv.org/abs/#1}{\nolinkurl{https://arxiv.org/abs/#1}}}

\bibitem[{Abell {et~al.}(2009)}]{LSSTScience:2009jmu}
Abell, P.~A., {et~al.} 2009.
\newblock \doarXiv{0912.0201}

\bibitem[{Aghanim {et~al.}(2018)}]{Aghanim:2018eyx}
Aghanim, N., {et~al.} 2018.
\newblock \doarXiv{1807.06209}

\bibitem[{Auger {et~al.}(2009)Auger, Treu, Bolton, Gavazzi, Koopmans, Marshall,
  Bundy, \& Moustakas}]{Auger:2009hj}
Auger, M.~W., Treu, T., Bolton, A.~S., {et~al.} 2009, Astrophys. J., 705, 1099,
  \dodoi{10.1088/0004-637X/705/2/1099}

\bibitem[{Auger {et~al.}(2010)Auger, Treu, Bolton, Gavazzi, Koopmans, Marshall,
  Moustakas, \& Burles}]{Auger:2010va}
---. 2010, Astrophys. J., 724, 511, \dodoi{10.1088/0004-637X/724/1/511}

\bibitem[{Bengochea \& Ferraro(2009)}]{Bengochea:2008gz}
Bengochea, G.~R., \& Ferraro, R. 2009, Phys. Rev. D, 79, 124019,
  \dodoi{10.1103/PhysRevD.79.124019}

\bibitem[{Bertotti {et~al.}(2003)Bertotti, Iess, \& Tortora}]{Bertotti:2003rm}
Bertotti, B., Iess, L., \& Tortora, P. 2003, Nature, 425, 374,
  \dodoi{10.1038/nature01997}

\bibitem[{Bolton {et~al.}(2008)Bolton, Burles, Koopmans, Treu, Gavazzi,
  Moustakas, Wayth, \& Schlegel}]{Bolton:2008xf}
Bolton, A.~S., Burles, S., Koopmans, L. V.~E., {et~al.} 2008, Astrophys. J.,
  682, 964, \dodoi{10.1086/589327}

\bibitem[{Bolton {et~al.}(2006{\natexlab{a}})Bolton, Burles, Koopmans, Treu, \&
  Moustakas}]{Bolton:2005nf}
Bolton, A.~S., Burles, S., Koopmans, L. V.~E., Treu, T., \& Moustakas, L.~A.
  2006{\natexlab{a}}, Astrophys. J., 638, 703, \dodoi{10.1086/498884}

\bibitem[{Bolton {et~al.}(2006{\natexlab{b}})Bolton, Rappaport, \&
  Burles}]{Bolton:2006yz}
Bolton, A.~S., Rappaport, S., \& Burles, S. 2006{\natexlab{b}}, Phys. Rev. D,
  74, 061501, \dodoi{10.1103/PhysRevD.74.061501}

\bibitem[{Bolton {et~al.}(2012)Bolton, Brownstein, Kochanek, Shu, Schlegel,
  Eisenstein, Wake, Connolly, Maraston, \& Weaver}]{Bolton:2012uh}
Bolton, A.~S., Brownstein, J.~R., Kochanek, C.~S., {et~al.} 2012, Astrophys.
  J., 757, 82, \dodoi{10.1088/0004-637X/757/1/82}

\bibitem[{Brownstein {et~al.}(2012)}]{Brownstein:2011leg}
Brownstein, J.~R., {et~al.} 2012, Astrophys. J., 744, 41,
  \dodoi{10.1088/0004-637X/744/1/41}

\bibitem[{Cao {et~al.}(2016)Cao, Biesiada, Yao, \& Zhu}]{Cao:2016wor}
Cao, S., Biesiada, M., Yao, M., \& Zhu, Z.-H. 2016, Mon. Not. Roy. Astron.
  Soc., 461, 2192, \dodoi{10.1093/mnras/stw932}

\bibitem[{Cao {et~al.}(2017)Cao, Li, Biesiada, Xu, Cai, \& Zhu}]{Cao:2017nnq}
Cao, S., Li, X., Biesiada, M., {et~al.} 2017, Astrophys. J., 835, 92,
  \dodoi{10.3847/1538-4357/835/1/92}

\bibitem[{Carroll(2001)}]{Carroll:2000fy}
Carroll, S.~M. 2001, Living Rev. Rel., 4, 1, \dodoi{10.12942/lrr-2001-1}

\bibitem[{Chen {et~al.}(2019)Chen, Li, Shu, \& Cao}]{Chen:2018jcf}
Chen, Y., Li, R., Shu, Y., \& Cao, X. 2019, Mon. Not. Roy. Astron. Soc., 488,
  3745, \dodoi{10.1093/mnras/stz1902}

\bibitem[{Collett {et~al.}(2018)Collett, Oldham, Smith, Auger, Westfall, Bacon,
  Nichol, Masters, Koyama, \& van~den Bosch}]{collett2018precise}
Collett, T.~E., Oldham, L.~J., Smith, R.~J., {et~al.} 2018, Science, 360, 1342

\bibitem[{Copeland {et~al.}(2006)Copeland, Sami, \&
  Tsujikawa}]{Copeland:2006wr}
Copeland, E.~J., Sami, M., \& Tsujikawa, S. 2006, Int. J. Mod. Phys. D, 15,
  1753, \dodoi{10.1142/S021827180600942X}

\bibitem[{Cui {et~al.}(2017)Cui, Li, \& Zhang}]{Cui:2017idf}
Cui, J.-L., Li, H.-L., \& Zhang, X. 2017, Sci. China Phys. Mech. Astron., 60,
  080411, \dodoi{10.1007/s11433-017-9054-5}

\bibitem[{Dainotti {et~al.}(2021)Dainotti, De~Simone, Schiavone, Montani,
  Rinaldi, \& Lambiase}]{Dainotti:2021pqg}
Dainotti, M.~G., De~Simone, B., Schiavone, T., {et~al.} 2021, Astrophys. J.,
  912, 150, \dodoi{10.3847/1538-4357/abeb73}

\bibitem[{Di~Valentino {et~al.}(2021)Di~Valentino, Mena, Pan, Visinelli, Yang,
  Melchiorri, Mota, Riess, \& Silk}]{DiValentino:2021izs}
Di~Valentino, E., Mena, O., Pan, S., {et~al.} 2021, Class. Quant. Grav., 38,
  153001, \dodoi{10.1088/1361-6382/ac086d}

\bibitem[{Dyson {et~al.}(1920)Dyson, Eddington, \& Davidson}]{Dyson:1920cwa}
Dyson, F.~W., Eddington, A.~S., \& Davidson, C. 1920, Phil. Trans. Roy. Soc.
  Lond. A, 220, 291, \dodoi{10.1098/rsta.1920.0009}

\bibitem[{Etherington(1933)}]{etherington1933lx}
Etherington, I. 1933, The London, Edinburgh, and Dublin Philosophical Magazine
  and Journal of Science, 15, 761

\bibitem[{Feng {et~al.}(2020)Feng, He, Li, Zhang, \& Zhang}]{Feng:2019jqa}
Feng, L., He, D.-Z., Li, H.-L., Zhang, J.-F., \& Zhang, X. 2020, Sci. China
  Phys. Mech. Astron., 63, 290404, \dodoi{10.1007/s11433-019-1511-8}

\bibitem[{Foreman-Mackey {et~al.}(2013)Foreman-Mackey, Hogg, Lang, \&
  Goodman}]{foreman2013emcee}
Foreman-Mackey, D., Hogg, D.~W., Lang, D., \& Goodman, J. 2013, Publications of
  the Astronomical Society of the Pacific, 125, 306

\bibitem[{Frieman {et~al.}(2013)Frieman, Collaboration,
  {et~al.}}]{frieman2013dark}
Frieman, J., Collaboration, D. E.~S., {et~al.} 2013, in American Astronomical
  Society Meeting Abstracts\# 221, Vol. 221, 335--01

\bibitem[{Gao {et~al.}(2021)Gao, Zhao, Xue, \& Zhang}]{Gao:2021xnk}
Gao, L.-Y., Zhao, Z.-W., Xue, S.-S., \& Zhang, X. 2021, JCAP, 07, 005,
  \dodoi{10.1088/1475-7516/2021/07/005}

\bibitem[{Geng {et~al.}(2015)Geng, Guo, He, Zhang, \& Zhang}]{Geng:2015hen}
Geng, J.-J., Guo, R.-Y., He, D.-Z., Zhang, J.-F., \& Zhang, X. 2015, Front.
  Phys. (Beijing), 10, 109501, \dodoi{10.1007/s11467-015-0507-8}

\bibitem[{Gerhard {et~al.}(2001)Gerhard, Kronawitter, Saglia, \&
  Bender}]{Gerhard:2000ck}
Gerhard, O., Kronawitter, A., Saglia, R.~P., \& Bender, R. 2001, Astron. J.,
  121, 1936, \dodoi{10.1086/319940}

\bibitem[{Guo {et~al.}(2019{\natexlab{a}})Guo, Zhang, \& Zhang}]{Guo:2018ans}
Guo, R.-Y., Zhang, J.-F., \& Zhang, X. 2019{\natexlab{a}}, JCAP, 02, 054,
  \dodoi{10.1088/1475-7516/2019/02/054}

\bibitem[{Guo {et~al.}(2020)Guo, Zhang, \& Zhang}]{Guo:2019dui}
---. 2020, Sci. China Phys. Mech. Astron., 63, 290406,
  \dodoi{10.1007/s11433-019-1514-0}

\bibitem[{Guo {et~al.}(2019{\natexlab{b}})Guo, Zhang, Zhang, \&
  Zhang}]{Guo:2018uic}
Guo, R.-Y., Zhang, L., Zhang, J.-F., \& Zhang, X. 2019{\natexlab{b}}, Sci.
  China Phys. Mech. Astron., 62, 30411, \dodoi{10.1007/s11433-018-9278-1}

\bibitem[{Guy {et~al.}(2010)}]{SNLS:2010pgl}
Guy, J., {et~al.} 2010, Astron. Astrophys., 523, A7,
  \dodoi{10.1051/0004-6361/201014468}

\bibitem[{Jiang \& Kochanek(2007)}]{Jiang:2007jx}
Jiang, G., \& Kochanek, C.~S. 2007, Astrophys. J., 671, 1568,
  \dodoi{10.1086/522580}

\bibitem[{Jorgensen {et~al.}(1995)Jorgensen, Franx, \&
  Kjaergaard}]{Jorgensen:1995zz}
Jorgensen, I., Franx, M., \& Kjaergaard, P. 1995, Mon. Not. Roy. Astron. Soc.,
  276, 1341

\bibitem[{Kessler \& Scolnic(2017)}]{Kessler:2016uwi}
Kessler, R., \& Scolnic, D. 2017, Astrophys. J., 836, 56,
  \dodoi{10.3847/1538-4357/836/1/56}

\bibitem[{Koopmans \& Treu(2002)}]{koopmans2002stellar}
Koopmans, L.~V., \& Treu, T. 2002, The Astrophysical Journal Letters, 568, L5

\bibitem[{Koopmans(2006)}]{Koopmans:2005ig}
Koopmans, L. V.~E. 2006, EAS Publ. Ser., 20, 161, \dodoi{10.1051/eas:2006064}

\bibitem[{Koopmans \& Treu(2003)}]{Koopmans:2002qh}
Koopmans, L. V.~E., \& Treu, T. 2003, Astrophys. J., 583, 606,
  \dodoi{10.1086/345423}

\bibitem[{Li {et~al.}(2015)Li, Zhang, \& Zhang}]{Li:2015poa}
Li, Y.-H., Zhang, J.-F., \& Zhang, X. 2015, Phys. Lett. B, 744, 213,
  \dodoi{10.1016/j.physletb.2015.03.063}

\bibitem[{Liao {et~al.}(2016)Liao, Li, Cao, Biesiada, Zheng, \&
  Zhu}]{Liao:2015uzb}
Liao, K., Li, Z., Cao, S., {et~al.} 2016, Astrophys. J., 822, 74,
  \dodoi{10.3847/0004-637X/822/2/74}

\bibitem[{Liao {et~al.}(2019)Liao, Shafieloo, Keeley, \&
  Linder}]{liao2019model}
Liao, K., Shafieloo, A., Keeley, R.~E., \& Linder, E.~V. 2019, The
  Astrophysical Journal Letters, 886, L23

\bibitem[{Liddle(2004)}]{Liddle:2004nh}
Liddle, A.~R. 2004, Mon. Not. Roy. Astron. Soc., 351, L49,
  \dodoi{10.1111/j.1365-2966.2004.08033.x}

\bibitem[{Mukherjee {et~al.}(1998)Mukherjee, Feigelson, Babu, Murtagh, Fraley,
  \& Raftery}]{Mukherjee:1998wp}
Mukherjee, S., Feigelson, E.~D., Babu, G.~J., {et~al.} 1998, Astrophys. J.,
  508, 314, \dodoi{10.1086/306386}

\bibitem[{Pan {et~al.}(2021)Pan, He, Qi, Li, Cao, Liu, \& Wang}]{Pan:2021tpk}
Pan, Y., He, Y., Qi, J., {et~al.} 2021, Astrophys. J., 911, 135,
  \dodoi{10.3847/1538-4357/abebe0}

\bibitem[{Pocino {et~al.}(2021)}]{Euclid:2021osj}
Pocino, A., {et~al.} 2021.
\newblock \doarXiv{2104.05698}

\bibitem[{Pound \& Rebka(1960)}]{Pound:1960zz}
Pound, R.~V., \& Rebka, Jr., G.~A. 1960, Phys. Rev. Lett., 4, 337,
  \dodoi{10.1103/PhysRevLett.4.337}

\bibitem[{Qi {et~al.}(2017)Qi, Cao, Biesiada, Zheng, \& Zhu}]{Qi:2017xzl}
Qi, J.-Z., Cao, S., Biesiada, M., Zheng, X., \& Zhu, H. 2017, Eur. Phys. J. C,
  77, 502, \dodoi{10.1140/epjc/s10052-017-5069-1}

\bibitem[{Qi {et~al.}(2019{\natexlab{a}})Qi, Cao, Zhang, Biesiada, Wu, \&
  Zhu}]{Qi:2018aio}
Qi, J.-Z., Cao, S., Zhang, S., {et~al.} 2019{\natexlab{a}}, Mon. Not. Roy.
  Astron. Soc., 483, 1104, \dodoi{10.1093/mnras/sty3175}

\bibitem[{Qi {et~al.}(2019{\natexlab{b}})Qi, Cao, Zheng, Pan, Li, Li, \&
  Liu}]{Qi:2019spg}
Qi, J.-Z., Cao, S., Zheng, C., {et~al.} 2019{\natexlab{b}}, Phys. Rev. D, 99,
  063507, \dodoi{10.1103/PhysRevD.99.063507}

\bibitem[{Qi {et~al.}(2016)Qi, Yang, Zhang, \& Liu}]{Qi:2014yxa}
Qi, J.-Z., Yang, R.-J., Zhang, M.-J., \& Liu, W.-B. 2016, Res. Astron.
  Astrophys., 16, 022, \dodoi{10.1088/1674-4527/16/2/022}

\bibitem[{Qi {et~al.}(2021)Qi, Zhao, Cao, Biesiada, \& Liu}]{Qi:2020rmm}
Qi, J.-Z., Zhao, J.-W., Cao, S., Biesiada, M., \& Liu, Y. 2021, Mon. Not. Roy.
  Astron. Soc., 503, 2179, \dodoi{10.1093/mnras/stab638}

\bibitem[{Reyes {et~al.}(2010)Reyes, Mandelbaum, Seljak, Baldauf, Gunn,
  Lombriser, \& Smith}]{Reyes:2010tr}
Reyes, R., Mandelbaum, R., Seljak, U., {et~al.} 2010, Nature, 464, 256,
  \dodoi{10.1038/nature08857}

\bibitem[{Riess {et~al.}(2019)Riess, Casertano, Yuan, Macri, \&
  Scolnic}]{Riess:2019cxk}
Riess, A.~G., Casertano, S., Yuan, W., Macri, L.~M., \& Scolnic, D. 2019,
  Astrophys. J., 876, 85, \dodoi{10.3847/1538-4357/ab1422}

\bibitem[{Ruff {et~al.}(2011)Ruff, Gavazzi, Marshall, Treu, Auger, \&
  Brault}]{Ruff:2010rv}
Ruff, A.~J., Gavazzi, R., Marshall, P.~J., {et~al.} 2011, Astrophys. J., 727,
  96, \dodoi{10.1088/0004-637X/727/2/96}

\bibitem[{Schwab {et~al.}(2010)Schwab, Bolton, \& Rappaport}]{Schwab:2009nz}
Schwab, J., Bolton, A.~S., \& Rappaport, S.~A. 2010, Astrophys. J., 708, 750,
  \dodoi{10.1088/0004-637X/708/1/750}

\bibitem[{Schwarz(1978)}]{Schwarz:1978tpv}
Schwarz, G. 1978, Annals Statist., 6, 461

\bibitem[{Scolnic {et~al.}(2018)}]{Scolnic:2017caz}
Scolnic, D.~M., {et~al.} 2018, Astrophys. J., 859, 101,
  \dodoi{10.3847/1538-4357/aab9bb}

\bibitem[{Seikel \& Clarkson(2013)}]{Seikel:2013fda}
Seikel, M., \& Clarkson, C. 2013.
\newblock \doarXiv{1311.6678}

\bibitem[{Seikel {et~al.}(2012{\natexlab{a}})Seikel, Clarkson, \&
  Smith}]{Seikel:2012uu}
Seikel, M., Clarkson, C., \& Smith, M. 2012{\natexlab{a}}, JCAP, 06, 036,
  \dodoi{10.1088/1475-7516/2012/06/036}

\bibitem[{Seikel {et~al.}(2012{\natexlab{b}})Seikel, Yahya, Maartens, \&
  Clarkson}]{Seikel:2012cs}
Seikel, M., Yahya, S., Maartens, R., \& Clarkson, C. 2012{\natexlab{b}}, Phys.
  Rev. D, 86, 083001, \dodoi{10.1103/PhysRevD.86.083001}

\bibitem[{Shapiro(1964)}]{Shapiro:1964uw}
Shapiro, I.~I. 1964, Phys. Rev. Lett., 13, 789,
  \dodoi{10.1103/PhysRevLett.13.789}

\bibitem[{Shu {et~al.}(2015)}]{Shu:2014aba}
Shu, Y., {et~al.} 2015, Astrophys. J., 803, 71,
  \dodoi{10.1088/0004-637X/803/2/71}

\bibitem[{Shu {et~al.}(2016{\natexlab{a}})Shu, Bolton, Kochanek, Oguri,
  P{\'{e}}rez-Fournon, Zheng, Mao, Montero-Dorta, Brownstein, Marques-Chaves,
  \& M{\'{e}}nard}]{Shu_2016}
Shu, Y., Bolton, A.~S., Kochanek, C.~S., {et~al.} 2016{\natexlab{a}}, The
  Astrophysical Journal, 824, 86, \dodoi{10.3847/0004-637x/824/2/86}

\bibitem[{Shu {et~al.}(2016{\natexlab{b}})Shu, Bolton, Mao, Kochanek,
  P{\'e}rez-Fournon, Oguri, Montero-Dorta, Cornachione, Marques-Chaves, Zheng,
  {et~al.}}]{shu2016boss}
Shu, Y., Bolton, A.~S., Mao, S., {et~al.} 2016{\natexlab{b}}, The Astrophysical
  Journal, 833, 264

\bibitem[{Shu {et~al.}(2017)}]{Shu:2017yon}
Shu, Y., {et~al.} 2017, Astrophys. J., 851, 48,
  \dodoi{10.3847/1538-4357/aa9794}

\bibitem[{Smith(2009)}]{Smith:2009fn}
Smith, T.~L. 2009.
\newblock \doarXiv{0907.4829}

\bibitem[{Song {et~al.}(2007)Song, Hu, \& Sawicki}]{Song:2006ej}
Song, Y.-S., Hu, W., \& Sawicki, I. 2007, Phys. Rev. D, 75, 044004,
  \dodoi{10.1103/PhysRevD.75.044004}

\bibitem[{Sonnenfeld {et~al.}(2013)Sonnenfeld, Treu, Gavazzi, Suyu, Marshall,
  Auger, \& Nipoti}]{Sonnenfeld:2013xga}
Sonnenfeld, A., Treu, T., Gavazzi, R., {et~al.} 2013, Astrophys. J., 777, 98,
  \dodoi{10.1088/0004-637X/777/2/98}

\bibitem[{Sonnenfeld {et~al.}(2015)Sonnenfeld, Treu, Marshall, Suyu, Gavazzi,
  Auger, \& Nipoti}]{Sonnenfeld:2014gpa}
Sonnenfeld, A., Treu, T., Marshall, P.~J., {et~al.} 2015, Astrophys. J., 800,
  94, \dodoi{10.1088/0004-637X/800/2/94}

\bibitem[{Sotiriou \& Faraoni(2010)}]{Sotiriou:2008rp}
Sotiriou, T.~P., \& Faraoni, V. 2010, Rev. Mod. Phys., 82, 451,
  \dodoi{10.1103/RevModPhys.82.451}

\bibitem[{Spergel {et~al.}(2015)}]{Spergel:2015sza}
Spergel, D., {et~al.} 2015.
\newblock \doarXiv{1503.03757}

\bibitem[{Taylor {et~al.}(1979)Taylor, Fowler, \& McCulloch}]{Taylor:1979zz}
Taylor, J.~H., Fowler, L.~A., \& McCulloch, P.~M. 1979, Nature, 277, 437,
  \dodoi{10.1038/277437a0}

\bibitem[{Thorne \& Will(1971)}]{Thorne:1971iat}
Thorne, K.~S., \& Will, C.~M. 1971, Astrophys. J., 163, 595,
  \dodoi{10.1086/150803}

\bibitem[{Treu \& Koopmans(2002)}]{Treu:2002ee}
Treu, T., \& Koopmans, L. 2002, Astrophys. J., 575, 87, \dodoi{10.1086/341216}

\bibitem[{Treu \& Koopmans(2004)}]{Treu:2004wt}
Treu, T., \& Koopmans, L. V.~E. 2004, Astrophys. J., 611, 739,
  \dodoi{10.1086/422245}

\bibitem[{Wang {et~al.}(2020)Wang, Qi, Zhang, \& Zhang}]{Wang:2019yob}
Wang, B., Qi, J.-Z., Zhang, J.-F., \& Zhang, X. 2020, Astrophys. J., 898, 100,
  \dodoi{10.3847/1538-4357/ab9b22}

\bibitem[{Wang {et~al.}(2021{\natexlab{a}})Wang, Ma, \& Xia}]{Wang:2020dbt}
Wang, G.-J., Ma, X.-J., \& Xia, J.-Q. 2021{\natexlab{a}}, Mon. Not. Roy.
  Astron. Soc., 501, 5714, \dodoi{10.1093/mnras/staa4044}

\bibitem[{Wang {et~al.}(2021{\natexlab{b}})Wang, He, Zhang, \&
  Zhang}]{Wang:2021kxc}
Wang, L.-F., He, D.-Z., Zhang, J.-F., \& Zhang, X. 2021{\natexlab{b}}.
\newblock \doarXiv{2102.09331}

\bibitem[{Weinberg(1989)}]{Weinberg:1988cp}
Weinberg, S. 1989, Rev. Mod. Phys., 61, 1, \dodoi{10.1103/RevModPhys.61.1}

\bibitem[{Yang {et~al.}(2020)Yang, Birrer, \& Hu}]{Yang:2020eoh}
Yang, T., Birrer, S., \& Hu, B. 2020, Mon. Not. Roy. Astron. Soc., 497, L56,
  \dodoi{10.1093/mnrasl/slaa107}

\bibitem[{Yang {et~al.}(2019)Yang, Hu, Cai, \& Wang}]{Yang:2018bdf}
Yang, T., Hu, B., Cai, R.-G., \& Wang, B. 2019, Astrophys. J., 880, 50,
  \dodoi{10.3847/1538-4357/ab271e}

\bibitem[{Zhang {et~al.}(2014)Zhang, Li, \& Zhang}]{Zhang:2014lfa}
Zhang, J.-F., Li, Y.-H., \& Zhang, X. 2014, Phys. Lett. B, 739, 102,
  \dodoi{10.1016/j.physletb.2014.10.044}

\bibitem[{Zhang {et~al.}(2007)Zhang, Liguori, Bean, \& Dodelson}]{Zhang:2007nk}
Zhang, P., Liguori, M., Bean, R., \& Dodelson, S. 2007, Phys. Rev. Lett., 99,
  141302, \dodoi{10.1103/PhysRevLett.99.141302}

\bibitem[{Zhang(2019)}]{Zhang:2019ylr}
Zhang, X. 2019, Sci. China Phys. Mech. Astron., 62, 110431,
  \dodoi{10.1007/s11433-019-9445-7}

\bibitem[{Zhao {et~al.}(2018)Zhao, Zhang, \& Zhang}]{Zhao:2017jma}
Zhao, M.-M., Zhang, J.-F., \& Zhang, X. 2018, Phys. Lett. B, 779, 473,
  \dodoi{10.1016/j.physletb.2018.02.042}

\end{thebibliography}
\bibliographystyle{aasjournal}

\end{document}